\documentclass[pre,amsmath,amssymb,floatfix,preprint,showpacs]{revtex4}
\usepackage{graphicx}
\usepackage{float}
\usepackage{bm}
\usepackage[utf8]{inputenc}

\begin{document}
\title{The effect of polydispersity on the ordering transition of adsorbed self-assembled rigid rods}
\author{N. G. Almarza}
\affiliation{Instituto de Qu{\'\i}mica F{\'\i}sica Rocasolano, CSIC, Serrano 119, E-28006 Madrid, Spain }
\author{J. M. Tavares}
\affiliation{Centro de F\'{\i}sica Te\'orica e Computacional, Universidade de Lisboa, Avenida Professor Gama Pinto 2,
P-1649-003 Lisbon, Portugal}
\affiliation{Instituto Superior de Engenharia de Lisboa, Rua Conselheiro Em\'{\i}dio Navarro 1, 
P-1950-062 Lisbon, Portugal}
\author{M. M. Telo da Gama}
\affiliation{Centro de F\'{\i}sica Te\'orica e Computacional, Universidade de Lisboa, Avenida Professor Gama Pinto 2,
P-1649-003 Lisbon, Portugal}
\affiliation{Departamento de F\'{\i}sica, Faculdade de Ci\^encias, Universidade de Lisboa, Campo Grande,
P-1749-016 Lisbon, Portugal}

\date{\today}
\begin{abstract}
Extensive Monte Carlo simulations were carried out to investigate the nature of the ordering
transition of a model of adsorbed self-assembled rigid rods on the bonds of a square lattice [Tavares et. al., 
Phys. Rev E {\bf 79}, 021505 (2009)]. The polydisperse rods undergo a continuous ordering transition that is found
to be in the two-dimensional Ising universality class, as in models where the rods are monodisperse. This finding 
is in sharp contrast with the recent claim that equilibrium polydispersity changes the nature of the phase transition 
in this class of models [L\'opez et. al., Phys. Rev E {\bf 80}, 040105(R)(2009)].  
\end{abstract}
\maketitle
%%%%%%%%%%%%%%%%%%%%%%%%%%%%%%%%%%%%%%%%%%%%%%%%%%%%%%%%%%%%%%%%%%%%%%%%%%%%%%%%%%%%%%%%%%%%
\section{Introduction}

It has been shown, recently, that pure hard-rod models in two-dimensions (2D) exhibit discrete orientational order 
without translational order \cite{Ioffe}, driven by a mechanism resembling that proposed by Onsager for the nematic 
transition of rods in three dimensions \cite{Onsager}. 
Specifically, it was proved that a system of rods on the square lattice, with hard-core exclusion and length 
distribution between 2 and $n$, exhibits discrete orientational long-range order for suitable fugacities  
and large $n$. This may seem surprising as the nature of the transition of monodisperse freely rotating rods 
in 2D remains subtle, as it appears to depend on the details of the particle interactions \cite{Straley,Frenkel,Vink}. 

Simple 2D restricted orientation models are relevant to describe the sub-monolayer regime of linear molecules 
adsorbed on crystalline substrates \cite{Potoff} and, even without polydispersity, rigid rod (RR) models were shown to 
exhibit a number of interesting features\cite{Matoz2008a,Matoz2008b,Ghosh}. It was found that the ordered phase is stable 
for sufficiently large aspect ratios\cite{Matoz2008a,Ghosh} and that the transition on the square lattice is 2D Ising \cite{Matoz2008b}.

Polydisperse restricted orientation RR models are generalizations of the Zwanzig model\cite{Zwanzig}, and provide a 
useful starting point for understanding the effects of polydispersity on the phase behavior of RRs\cite{Clarke}.
The description of self-assembled rods has to consider not only the effects of polydispersity but also the polymerization process. 
In this context, we proposed a model of self-assembled RR (SARR), composed of monomers with two bonding sites that polymerize 
reversibly into polydisperse chains \cite{Tavares2009a}. 
In the (lattice) model a site can be either occupied or unoccupied and each occupied site has a spin variable. On the square lattice,
the spins take two values representing the discretised set of orientations of the bonding sites that coincide with the lattice bonds. 
The interaction between two spins depends not only on their relative orientations but also on their orientations relative to the lattice 
bond connecting the monomers. We used a simple theory to investigate the interplay between self-assembly and ordering over the full 
range of temperature and density. The results revealed that the continuous ordering transition is predicted semiquantitatively by the 
theory\cite{Tavares2009a}. The universality class of this transition was not investigated; ordering of SARRs was assumed to be that of 
monodisperse rigid rods, which was found to be 2D Ising on this lattice\cite{Matoz2008b}.
 
In fact, the transition of polydisperse RRs, on the square lattice, was investigated for a vertex model that allows configurations 
promoting the polymerization of rods, in such a way that it is equivalent to the hard square model on the diagonal lattice. In polymer 
language, the ordered phase is stable when the average polymer length is long or its density is high. Calculations of the order-parameter 
using a variant of the density matrix renormalization group exhibit clear 2D Ising exponents ($\beta=0.125$) at all densities\cite{Takasaki}.
However, in 2009, Lopez et. al \cite{Lopez} carried out Monte Carlo (MC) simulations to investigate the critical behavior of the SARR model 
and concluded that self-assembly affects the nature of the transition, claiming it to be in the q=1 Potts class (random percolation), rather 
than in the 2D Ising (q=2 Potts). This is at odds with exact results\cite{Ioffe} that map the polydisperse RR model, with $n = \infty$, to 
the 2D Ising model, as well as with the results of the vertex model \cite{Takasaki} referred to above.

Apart from its fundamental interest, self-assembly is a very active field of research, driven by the goal of designing 
new functional materials, inspired by biological processes where it is used routinely to construct robust supramolecular 
structures. In this context, the effect of polydispersity on the nature of the ordering transition of a given model is 
an important open question. 

In the following, we report the results of a systematic investigation of the criticality of the SARR model over a wide 
range of temperatures, corresponding to critical densities that decrease from 1.0 (full lattice) to 0.1. We note that the full lattice 
SARR model may be mapped to a 2D Ising model, while the zero density SARR model exhibits an equilibrium polymerization transition at zero 
temperature. The results of our simulations provide strong evidence that the transition remains in the 2D Ising class at all 
(finite) densities. However, the numerical results also suggest that the scaling region is strongly affected by the density, decreasing 
as the density decreases, in a way that depends both on the scaling variable and on the thermodynamic function under investigation.  

This paper is organized as follows: the model and the simulation methods are described in Sec. II. The results for 2D Ising criticality 
of the SARR model are reported in section III. In section IV we give additional arguments that support our conclusion: (i) we map the full 
lattice limit (FLL) onto the 2D Ising model, (ii) we consider the zero density limit and estimate the crossover line from the zero density 
'equilibrium polymerization transition' and (iii) we discuss the non-monotonic behavior of the internal energy per particle on the critical 
line. Finally, in section V we summarize our results, and offer an explanation for the q=1 Potts behavior observed by L\'opez et al.\cite{Lopez}.

\section{The model}
The model is the two bonding site model, on the square lattice, proposed in Ref. \cite{Tavares2009a} in the context of a 
general framework to understand self-assembly (see \cite{Tavares2009b,Tavares2010} and references therein). A lattice site is either empty 
or occupied by one monomer with two bonding sites. Each monomer, $i$, adopts one of two orientations, ${\bf s}_i= {\hat x}$ or ${\bf s}_i ={\hat y}$, corresponding to the alignment of the bonding sites with the lattice directions, $\hat x$ and $\hat y$. Monomers attract each other if their 
bonding sites overlap, promoting the self-assembly of polydisperse rigid rods. The energy of the system may be written as:
\begin{equation}
U = - \epsilon \sum_{i=1}^M \sum_{{\hat \alpha}={\hat  x},{\hat  y}} 
|{{\bf s}}({\bf r}_i) \cdot 
{{\bf s}}({\bf r}_i+{\hat \alpha})| 
 |{\bf s}({\bf r}_i) \cdot  {\hat \alpha}|,
\label{uij} 
\end{equation}
where $i$ labels a lattice site, ${\bf s}({\bf r})$ denotes the monomer orientation 
(${\bf s} = {\bf 0}$ for an empty site); ${\hat x}$ and ${\hat y}$ are lattice unit vectors, 
and $M$ is the total number of sites. 

The criticality of this model was investigated in Ref. 
\cite{Lopez} where it was found that polydispersity changes the nature of the ordering transition.  
In order to check this claim we have studied the model over a wide range of thermodynamic parameters, using a 
multicanonical MC method based on a Wang-Landau sampling scheme. We considered systems with sizes $L_x=L_y=L$, $M=L^2$ sites, 
and periodic boundary conditions (PBC). The simulation methods were used in previous studies and details may be found 
there\cite{Hoye,Almarza,Lomba}. Briefly, in a simulation run we fix the the system size, $L$, and the temperature $T$; we 
sample over the number of particles $0\le N \le M$ and attempt exclusively MC moves of insertion and deletion (with equal 
probability).
In the insertion attempts the orientation of the particle is chosen at random. The probability of a configuration, 
${\bf R}^{N}$, with $N$ particles is:
\begin{equation}
P({\bf R}^N|M,T) \propto w (N)  \exp \left[ - {\cal U} ( {\bf R}^{N}) /k_B T  \right]
\label{prn}
\end{equation}
where $k_B$ is the Boltzmann constant and the function $w(N)$ is chosen to ensure uniform sampling of the density. 
The probability of a configuration with $N$ particles is:
\begin{equation}
P(N|T) \propto w (N)  \int {\textrm d} {\bf R}^N \exp \left[ - {\cal U} ( {\bf R}^{N}) /k_B T  \right]
= w(N) e^{-A(N,M,T)/k_B T},
\end{equation}
where $A(N,M,T)$ is the Helmholtz free energy. The weight function required for uniform sampling of $N$, in the range 
$[0,M]$, satisfies:
\begin{equation}
w(N) \simeq  e^{A(N,M,T)/k_B T}/(M+1).
\end{equation}
Clearly, the Helmholtz free energy $A(N,M,T)$ is not known a priori, but appropriate estimates of $w(N)$
may be obtained \cite{Almarza} using a Wang-Landau-like method\cite{Wang}. The multi-canonical simulation and the 
computation of the required observables (energy, order parameters, etc.) are then carried out for $0\le N \le M$.
In line with previous work, we define the order parameter as\cite{Tavares2009a,Lopez}:
\begin{equation}
\delta = \frac{|N_x-N_y|}{N},
\end{equation}
where $N_x$ and $N_y$ are the number of monomers oriented in the directions $x$ and $y$, respectively.

The ordering transition, at a given temperature, is located by searching for pseudo-critical values
of the chemical potential, $\mu_c(L,T)$. We note that L\'opez et al.\cite{Lopez} used the density, $\rho=N/M$, as 
the control parameter. In the SARR model at fixed $T$, the chemical potential $\mu$ is the only external field and 
plays the role of the temperature $T$ in standard (full lattice) Potts simulations\cite{Landau_Binder}. 
We proceed by defining analogues of the Ising response functions, related with the second derivatives of
$\Phi/k_B T$ (with $\Phi = A - N \mu$, the Grand Potential) with respect to the {\it coupling constant}
$K = 1/k_B T$, and $\mu$:
\begin{equation}
c =  - \frac{1}{k_B T^2 V } 
\frac{ \partial^2 \left[ K \Phi(\mu,M,K) \right] }{\partial K^2 } =
\left( \frac {\partial \left[ {\bar u} - \mu \rho \right]  } {\partial T }\right)_{\mu,M};
\end{equation}
\begin{equation}
\rho'_{\mu} = - \frac{ k_B T}{V} 
\frac{ \partial^2 \left[ K \Phi(\mu,M,K) \right] }{\partial \mu^2 } =
\left( \frac {\partial   \rho  } {\partial \mu }\right)_{T,M},
\end{equation}
where ${\bar u} \equiv U/M$. The quantities $c$, and $\rho'_{\mu}$ are expected
to scale at the critical point as\cite{Landau_Binder}:
\begin{equation}
c(L,\mu_c(T))  \sim L^{\alpha/\nu},
\end{equation}
\begin{equation}
\rho'_{\mu}(L,\mu_c(T))  \sim L^{\alpha/\nu},
\end{equation}
where $\alpha$, and $\nu$ are the specific heat and correlation length critical exponents.

We carried out MC simulations at several temperatures. At each temperature, a range of system 
sizes was considered; up to $L=144$ at reduced temperatures, $T^*\equiv k_BT/\epsilon \le0.25$ 
and up to $L=112$ at higher
temperatures. The results of each simulation are used to calculate histograms of the 
different observables that were then computed in terms of the chemical potential.

\section{Results}

\subsection{Binder cumulant}
\begin{figure}
\includegraphics[width=80mm,clip=]{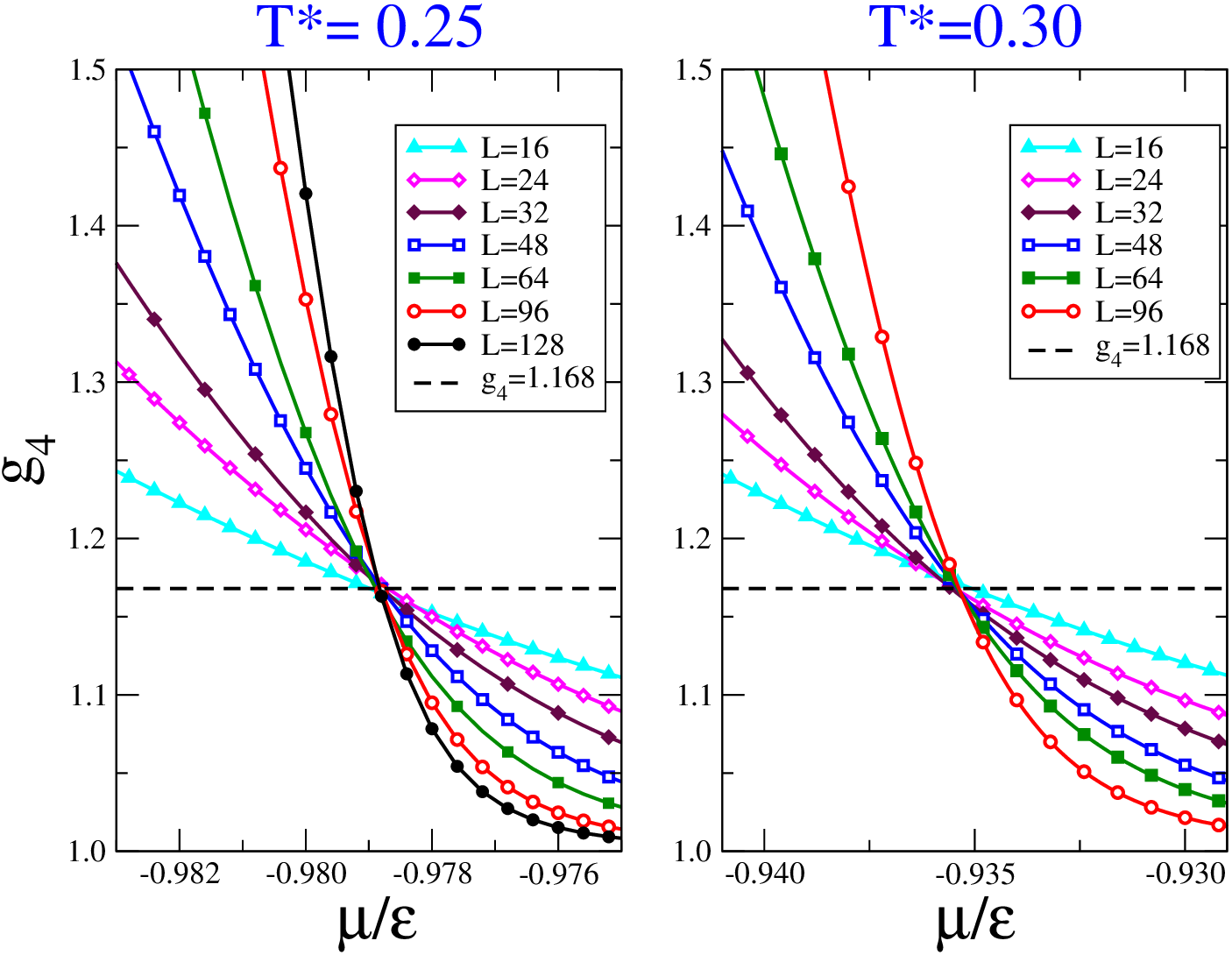}
\caption{(Color on line) Fourth order Binder cumulant as a function of $\mu$ for different system sizes, at
$T^*=0.25$ and $T^*=0.30$.}
\label{Fig1}
\end{figure}

We start by computing the fourth-order Binder cumulant\cite{Landau_Binder}, 
\begin{equation}
g_4 = \frac {< \delta^4 >}{<\delta^2>^2 },
\end{equation}
as a function of the chemical potential. In Fig. \ref{Fig1} we plot
$g_4(\mu)$ for different system sizes, at $T^*=0.25$ and $T^*=0.30$. 
It is clear that the cumulants for different system sizes, $L$, cross at a 
value of $g_4$ that is very close to the universal critical value for 2D Ising systems, with PBC 
 and $L_x=L_y$ (i.e. $g_4^{c}\simeq 1.168 $)\cite{Salas}. This immediately suggests 
that the criticality of the SARR model is in the 2D Ising (q=2 Potts) class (Q2UC), in  
contrast with the findings of L\'opez et al.\cite{Lopez}. Similar results were obtained for 
all the other temperatures investigated. 

\subsection{Computation of $\left( \partial \rho / \partial \mu \right)_T$ }

\begin{figure}
\includegraphics[width=80mm,clip=]{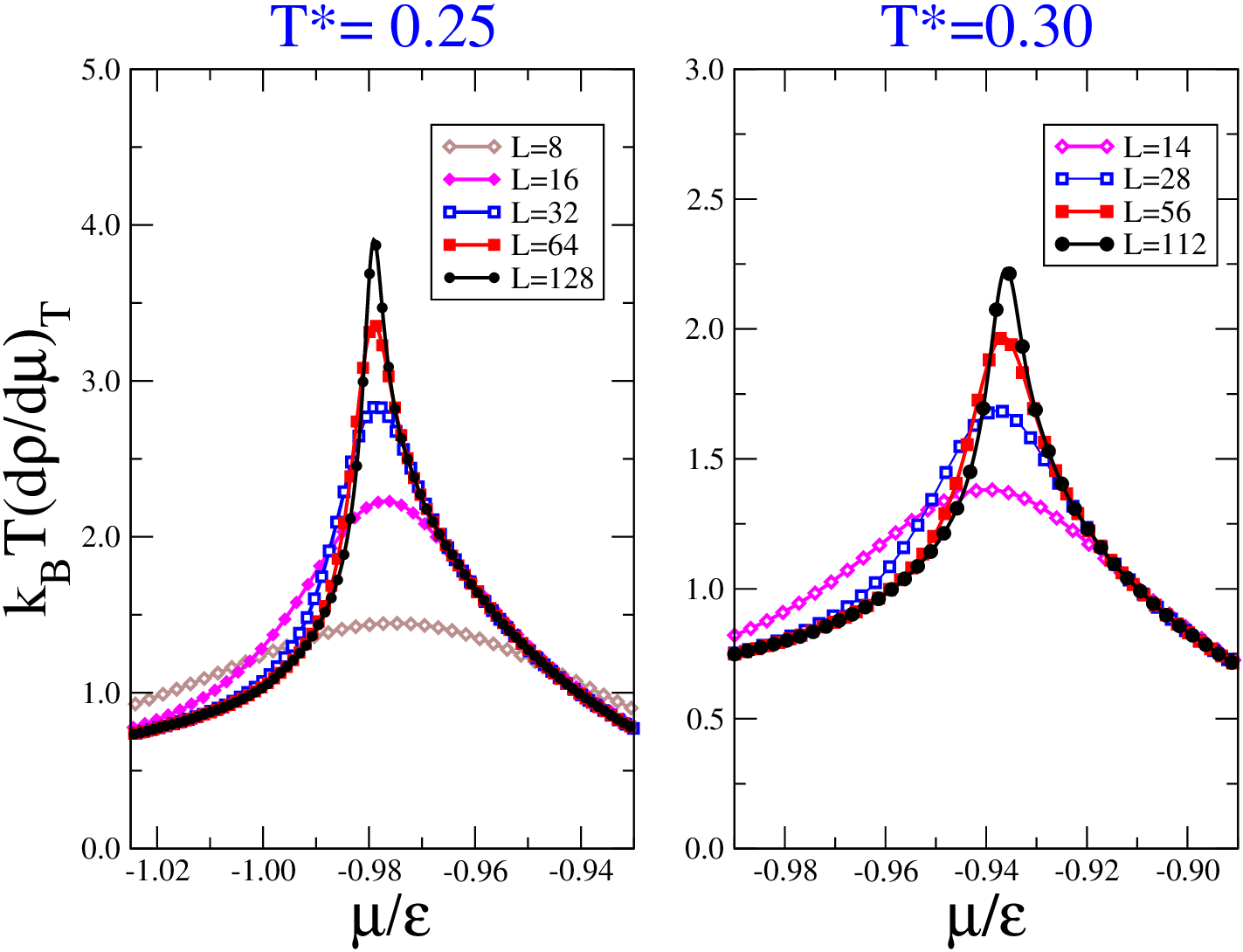}
\caption{(Color on line) Derivative of the density, $\rho$, as a function of the chemical potential, $\mu$, for different 
system sizes, at $T^*=0.25$ and $T^*=0.30$. A singularity is clearly signalled at both temperatures.}
\label{Fig2}
\end{figure}

\begin{figure}
\includegraphics[width=80mm,clip=]{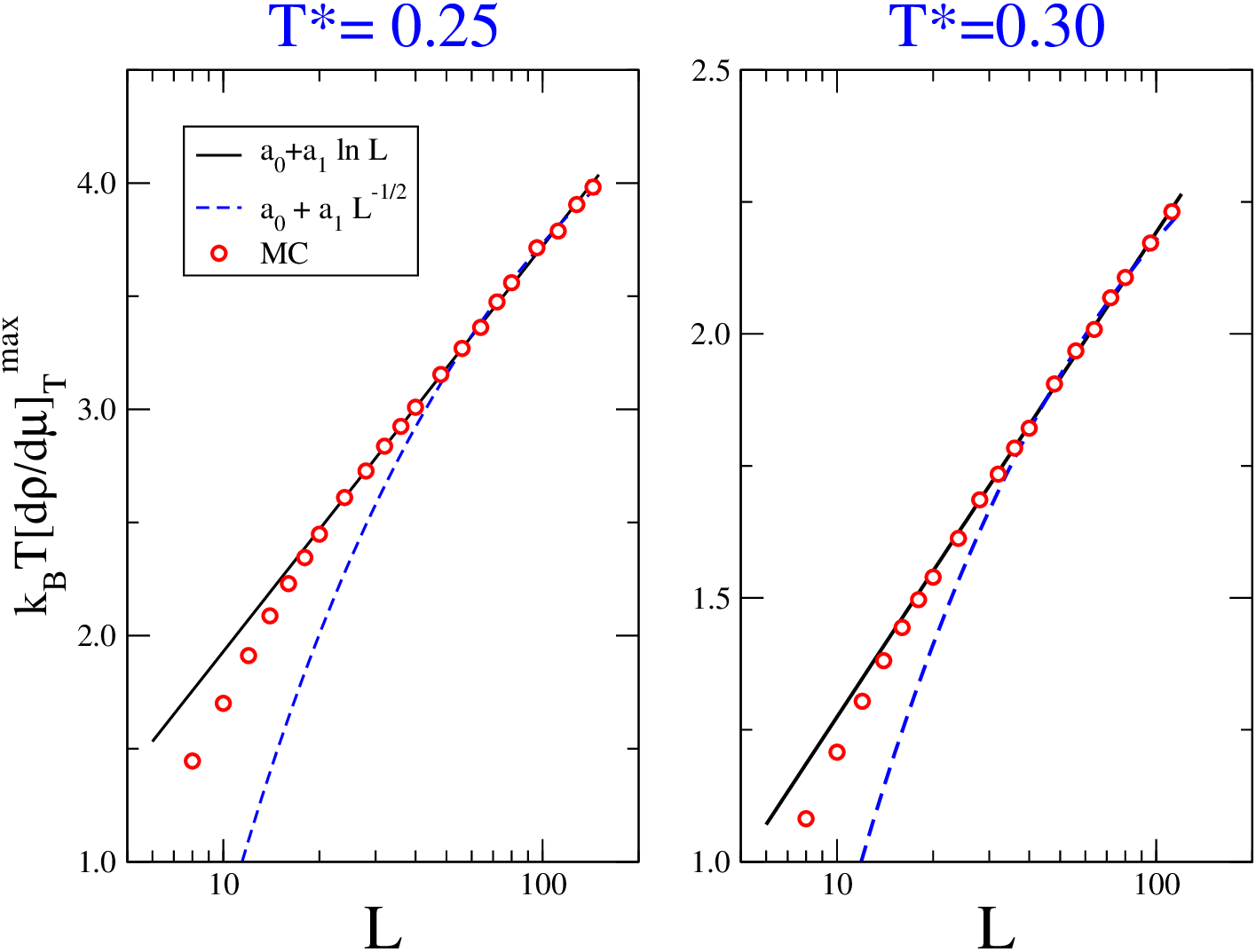}
\caption{System size dependence of the maximum of $k_BT (\partial \rho/\partial \mu)_T$, 
at $T^*=0.25$ and $T^*=0.30$. Simulation results are represented by points. Lines are 
fits to the scaling laws discussed in the text.}
\label{Fig3}
\end{figure}

In Fig. \ref{Fig2} we plot the derivative of the density with respect to the chemical potential, 
$\rho'_{\mu}$, as a function of $\mu$, for different system sizes, at the same temperatures $T^*=0.25$ and 
$T^*=0.30$. 
At values of the chemical potential, $\mu$, close to its critical value, the derivative of 
the density, $\rho'_{\mu}$, exhibits clear signs of 
singular behavior, with a peak that increases as the system size increases. 
The size dependence of the peaks is analyzed in Fig. \ref{Fig3}. 
The scaling of the peaks of $(\rho'_{\mu})^{max}(L,T)$, with the system size $L$, 
is characteristic of the universality class of the transition\cite{Landau_Binder}. For SARR on the 
square lattice we anticipate either q=1 Potts (Q1UC) behavior ($\alpha/\nu=-1/2$) \cite{Wu} as reported 
in Ref. \cite{Lopez} or Q2UC behavior, as found for monodisperse rods on the same 
lattice\cite{Matoz2008b}. In the latter case $\alpha/\nu=0$ and the peak is expected to diverge logarithmically\cite{Wu}.
The two scaling laws are:
\begin{equation}
(\rho'_{\mu})^{max}(L) =
a_0 + a_1 L^{-1/2} ,  \textrm{ for Q1UC}, 
\label{ddq1}
\end{equation}
\begin{equation}
(\rho'_{\mu})^{max}(L) = 
a_0 + a_1 \ln L ,  \textrm{ for Q2UC}. 
\label{ddq2}
\end{equation}
In Fig.
\ref{Fig3}, we plot fits of the two scaling laws to the simulation data. In both cases we discarded the data for the 
smallest system sizes. We found that the simulation results are better described by the 2D Ising scaling law, at 
all temperatures. In fact, equation (\ref{ddq2}) fits the data over a broader range of system sizes ($L \ge 24$ at 
$T^*=0.25$ and  $L \ge 32 $  at $T^*=0.30$) than does Eq. \ref{ddq1} ($L \ge 56$ at both temperatures).

\subsection{Critical Line}
\begin{figure}
\includegraphics[width=80mm,clip=]{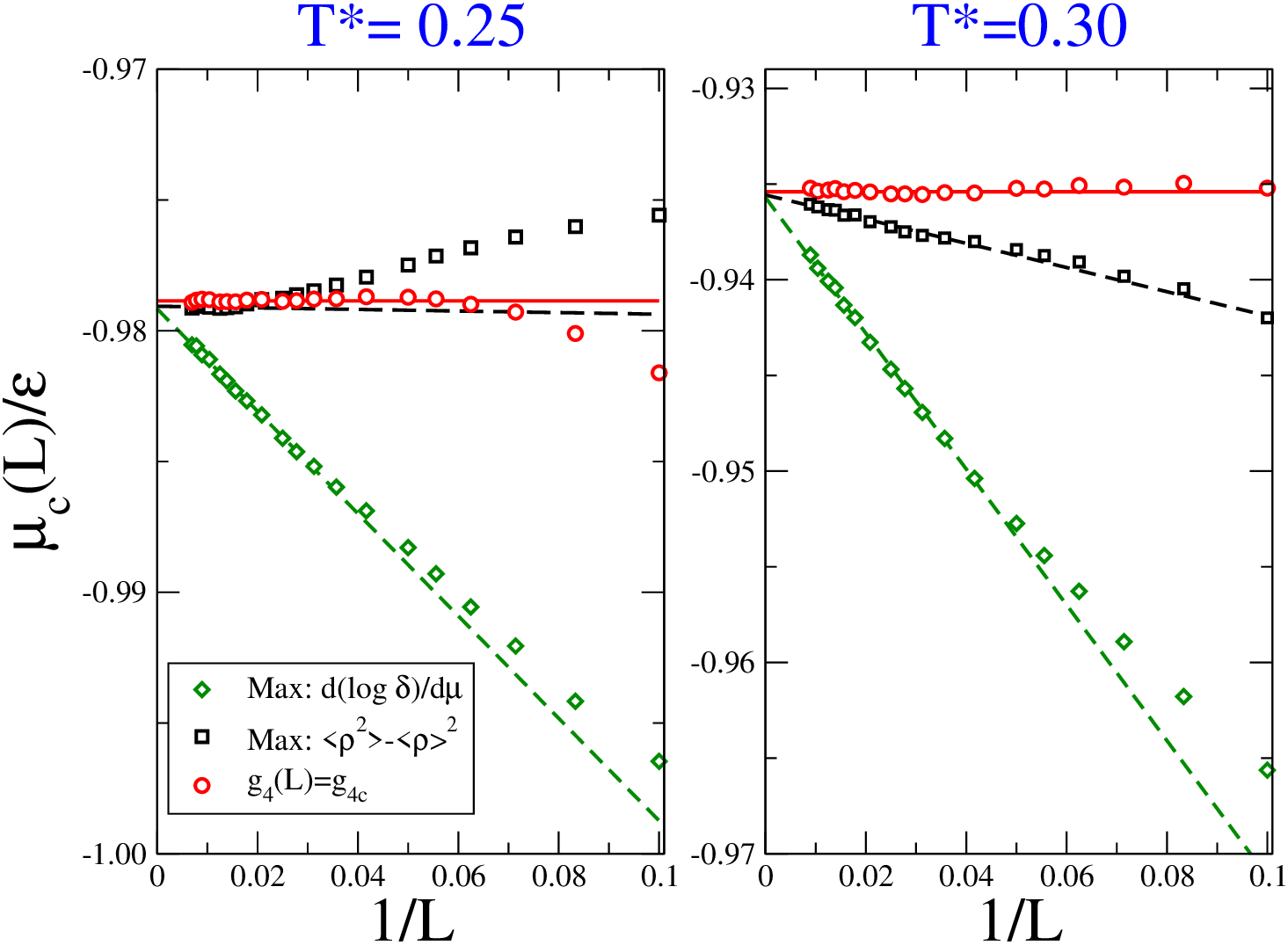}
\caption{(Color on line) System size dependence of the pseudo-critical chemical potentials. See the text and the 
legend for details. Lines are least-squares fits to the MC results.}
\label{Fig4}
\end{figure}

\begin{figure}[h]
\includegraphics[width=100mm,clip=]{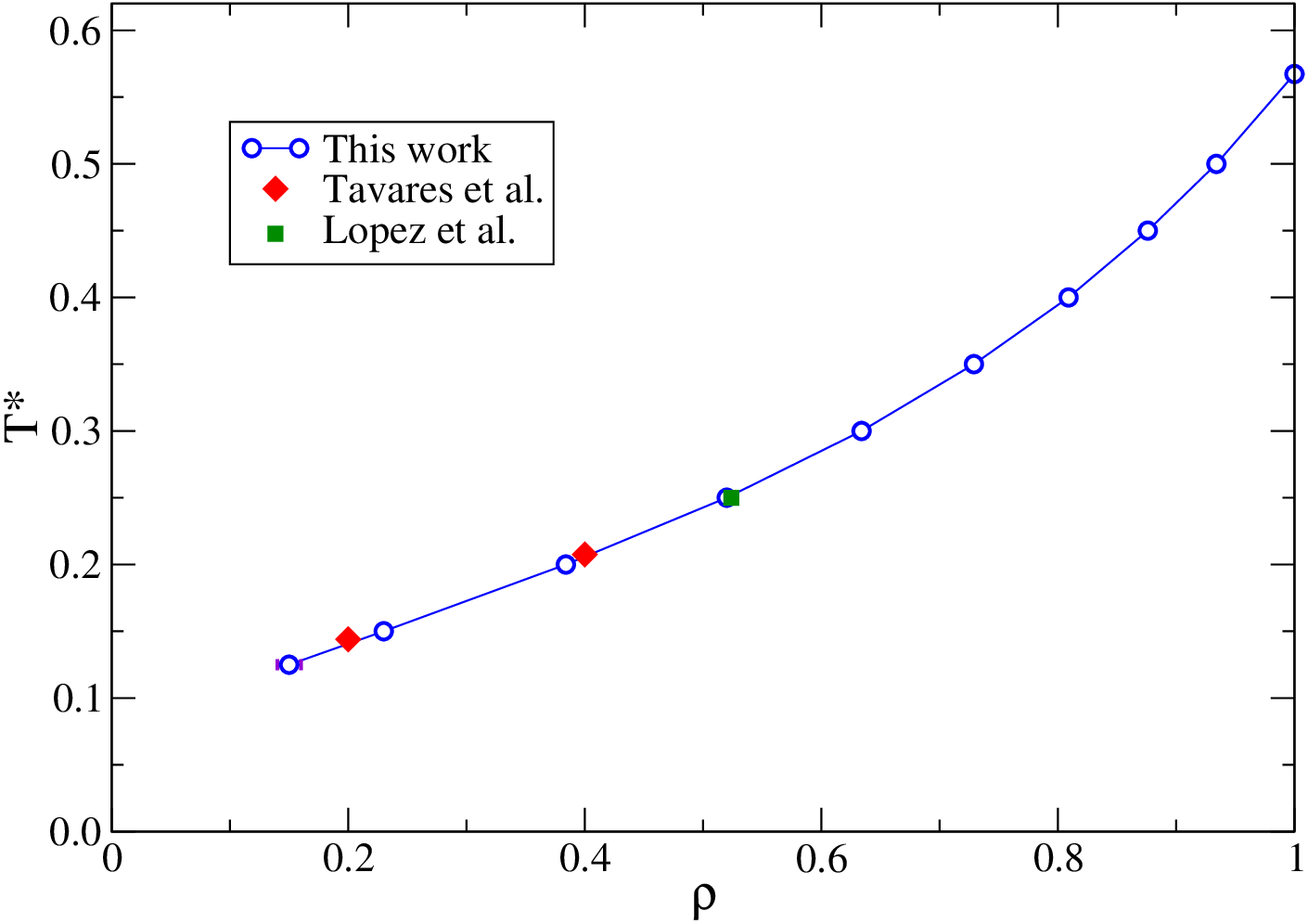}
\caption{(Color on line) Critical line of the SARR model. The diamonds and square are from Refs. [\onlinecite{Tavares2009a}],
and [\onlinecite{Lopez}], respectively.}
\label{fig.phase_diagram}
\end{figure}

We start by defining the pseudo-critical chemical potentials at fixed temperature. We 
consider the Binder cumulants, as functions of the chemical potential, and define the 
pseudo-critical chemical potentials, $\mu_{c}^{(L)}\equiv \mu_c(L)$ (at given $T$), such that: 
\begin{equation}
g_4(L,\mu_{c}^{(L)},T) = g_4^c.
\label{mucg4}
\end{equation}
We have also used different definitions based on the position of the maxima of the density 
fluctuations and of $(\partial \ln <\delta>/\partial \mu)_T$, to check the consistency 
of the results.
In Fig. \ref{Fig4} we plot the pseudo-critical chemical potentials as functions of $1/L$,
at $T^*=0.25$ and $T^*=0.30$. At both temperatures, and $L\ge 32$, the chemical potential, 
$\mu_c(L)$, computed using Eq. (\ref{mucg4}) is almost independent of system size (horizontal 
lines in the left and right panels of Fig. \ref{Fig4}). 
Estimates of the critical chemical potential, $\mu_c$, obtained by extrapolating the MC results, are collected in 
Table \ref{table.tc}. 
Considering the behavior of $\mu_c(L)$ obtained using Eq. (\ref{mucg4}), we used the results 
of simulations at this chemical potential, to compute the critical density $\rho_c$ 
and the critical exponents. Assuming 2D Ising behavior, the critical density is given 
by\cite{Landau_Binder}:
\begin{equation}
\rho_c(L,\mu_c^{(L)},T) = \rho_c(T) + a L^{-1}.
\end{equation}
The results for the critical line $T_c(\rho)$ are plotted in Fig. \ref{fig.phase_diagram}. As 
expected, the temperature at the ordering transition decreases as the density decreases. The critical
points calculated in earlier work (diamonds[\onlinecite{Tavares2009a}] and square[\onlinecite{Lopez}]) 
fall on the critical line, within the statistical error (open circles). The line of critical points of the
SARR model continues beyond the lowest density reported in Fig. \ref{fig.phase_diagram}. However, the 
rapid increase of the average length of the rods at these (low) densities and temperatures prevents  
an efficient simulation of these systems with the currently available techniques.    
\subsection{Critical Exponents}
The critical exponents $\beta/\nu$, and $\gamma/\nu$ \cite{Lopez,Landau_Binder} 
were estimated by fitting the MC results to the scaling laws,
\begin{equation}
\log \delta(L,\mu_c^{(L)},T) =  a_{\beta} - \frac{\beta}{\nu} \log L,
\label{eq.betanu}
\end{equation}
and
\begin{equation}
\log \chi(L,\mu_c^{(L)},T) =  a_{\xi} + \frac{\gamma}{\nu} \log L,
\end{equation}
where the susceptibility $\chi$ is defined as:
\begin{equation}
\chi = L^2 \left[ <\delta^2 > - < \delta >^2 \right] /k_B T.
\end{equation}
\begin{table}[h]
\begin{tabular}{|c|ccccc|}
\hline \hline
$T^*$ &  $\mu_c/\epsilon $ & $\rho_c$ & $<U/N\epsilon>_c$ & $\beta/\nu$ & $\gamma/\nu$   \\ 
\hline 
0.125 & -1.0030(5) & 0.15(1) & -0.944(1) & 0.08(2) & 1.86(3) \\
0.15 & -1.0038(1) & 0.230(3) & -0.919(1) &  0.10(2) & 1.79(4)  \\
0.20 & -0.9991(1) & 0.384(1) & -0.874(1)  & 0.108(4) & 1.761(9)  \\
0.25 & -0.9789(2) & 0.520(1) &  -0.843(1) & 0.110(2) & 1.757(4)  \\
0.30 & -0.9354(2) & 0.634(1) &  -0.826(1) & 0.111(2) & 1.754(3) \\
0.35 & -0.8597(6) & 0.729(1) & -0.818(1) & 0.114(2) & 1.750(2)  \\
0.40 & -0.7383(3) & 0.8086(4) & -0.8188(3) & 0.116(2) & 1.751(3)  \\
0.45  & -0.5452(5) & 0.8760(2) & -0.8249(3) & 0.120(3) & 1.749(3)  \\
0.50   & -0.214(2) & 0.9338(4) & -0.8345(5) & 0.123(2) & 1.751(4)  \\
0.5673   &   $\infty$ & 1.000    & -0.85355  & 0.125    & 1.750     \\
\hline 
\hline 
\end{tabular}
\caption{Results for the critical parameters and effective critical exponents.}
\label{table.tc}
\end{table}
The critical parameters and effective exponents are collected in Table \ref{table.tc} 
at 10 different temperatures.
Note that the exponents computed for $\beta/\nu$ lie between those corresponding to 
the Q1UC ($\beta/\nu=5/48\simeq 0.104$)\cite{Lopez,Wu} at low temperatures and those 
corresponding to the Q2UC ($\beta/\nu = 1/8$)\cite{Wu} at high temperatures.
The results for $\gamma/\nu$ are closer to those of Q2UC ($\gamma/\nu=7/4$) over a 
wider range of temperature but at the lowest temperatures they also approach those 
of Q1UC ($\gamma/\nu=43/24\simeq 1.792$).  

We note that Eq. (\ref{eq.betanu}) does not provide a good fit of the 
simulation results for a wide range of system sizes, and 
thus the results for $\beta/\nu$ should be regarded as effective exponents.
Consideration of higher-order finite size corrections, of the form, $\delta_c(L) = L^{-\beta/\nu} \left[ a_0 + b L^{-\omega} \right]$,
is not a feasible, as fits of the simulation results with four adjustable parameters 
cannot discriminate between these two, similar, scaling laws.  

\begin{figure}[h]
\includegraphics[width=100mm,clip=]{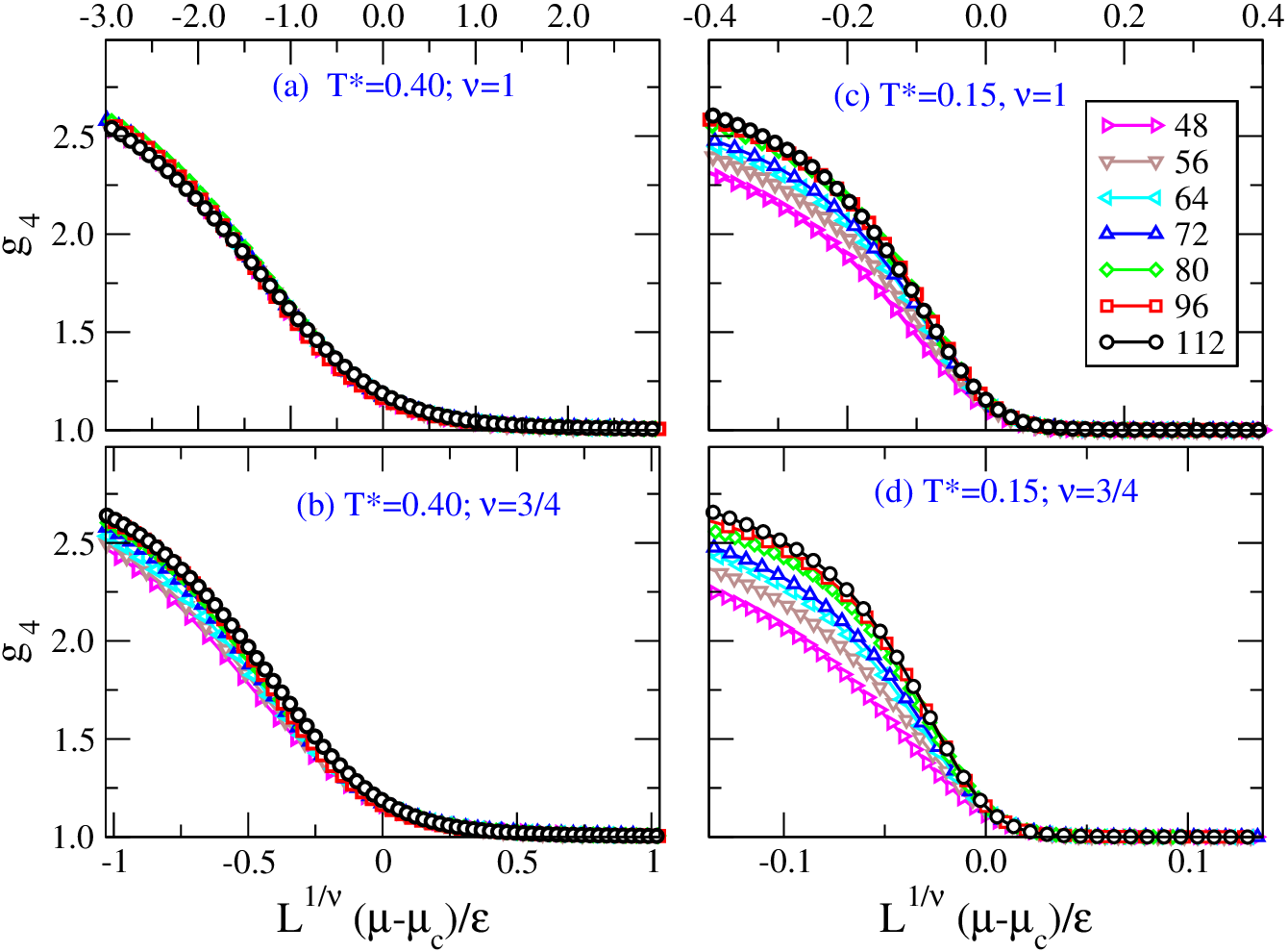}
\caption{(Color on line) Fourth order Binder cumulant as a function of $\mu$ for different system sizes, 
at $T^*=0.15$ and $T^*=0.40$.}
\label{Fig6}
\end{figure}

To proceed, we have investigated the finite-size scaling of the Binder cumulant, at 
$T^*=0.15$ and $T^*=0.40$. In Fig. \ref{Fig6} we plot $g_4(\mu)$ versus 
$L^{1/\nu}(\mu-\mu_c)$, using for the critical exponent, $\nu$, the values corresponding 
to the q=1 ($\nu=4/3$) and q=2 ($\nu=1)$ universality classes.
At $T^*=0.40$, the data collapse with $\nu=1$ is excellent, confirming that scaling 
for the Q2UC is satisfied for all the system sizes ($L\ge 48$). However, at the lowest 
temperature, $T^*=0.15$, the data collapse fails for both universality classes and systems 
with $L<80$. Nevertheless, the collapse observed with the Q2UC exponents is 
marginally better than that observed with the Q1UC exponents, suggesting that the SARR criticality 
is still in the 2D Ising class.

Finally, we analyzed the scaling behavior of the derivative of the logarithm of the order 
parameter with respect to the chemical potential, at constant temperature. The maxima
of this quantity are expected to scale with the system size as \cite{Ferrenberg,Landau_Binder,Lopez}
$Q'_{\mu} = \left( \partial \ln \delta /\partial \mu \right)_{T,L}^{\textrm max} \sim L^{1/\nu}$.
On the basis of this scaling law, we have computed effective values of $1/\nu$ by Chi-square 
fitting\cite{Numerical_Recipes} the simulation
results to $ Q'_{\mu}(L) = a L^{1/\nu}$.
The results for $1/\nu$ are collected in table \ref{table.nu}, for several temperatures, and confirm that the 
SARR model is in the 2D Ising class. It is also clear that as the temperature decreases the effect of the finite 
system size becomes more important, i.e. one requires larger systems to stay on the asymptotic scaling region.
\begin{table}[h]
\begin{tabular}{|c|ccccc|}
\hline \hline
$T^*$ &    $n$    & $ L_{min}$  &  $  L_{max} $& $ \chi^2/{\textrm  d.o.f } $ & $ 1/\nu $ \\
\hline
0.15  &  5   & 80  &        144  &     0.84     &       1.22(15) \\
0.20  & 5    & 80  &      144    &  1.71        &    1.13(10) \\
0.25  & 5    & 80  &       144   &   1.13       &     1.04(9) \\
0.30  & 8    & 48  &       112   &   0.08       &     1.05(4) \\
0.35  & 8    & 40  &       112   &   0.65       &     1.02(2) \\
0.40  &  10  & 32  &       112   &   1.33       &     1.00(2) \\
\hline
\hline 
\end{tabular}
\caption{Estimates of effective values of $1/\nu$ for different temperatures;
$n$ is the number of points (system sizes) used in the fitting; $L_{min}$, 
and $L_{max}$ are the minimum and maximum system sizes considered, with $L_{min}$
chosen to provide statistically acceptable values for the $\chi^2$ merit function
\cite{Numerical_Recipes}. d.o.f. is the number of degrees of freedom in each
fitting.}
\label{table.nu}
\end{table}

\section{Analysis of the criticality of the SARR model}

\subsection{The full lattice limit: 2D Ising}

The results of the previous section suggest clearly that the criticality of the SARR model is 
in the 2D Ising class. In this section we investigate the full lattice limit of the SARR model, or full 
lattice limit (FLL) for short, where we can prove that this is indeed the case. One also expects the 
criticality to remain unchanged, as long as no other transitions occur on the critical line\cite{Silva}.

We performed a number of simulations using a multi-temperature algorithm 
proposed by Zhang and Ma\cite{Zhang} and found that the critical temperature is the 
same as that of the lattice gas. This is more than a coincidence as shown below.

We have mapped the monomer orientations ${\hat x}$, ${\hat y}$ to the Ising spins $\pm 1$ and computed 
the total energy of the models by adding the contributions, $u_p$, of square elementary plaquettes,
\begin{equation}
U = \frac{1}{2} \sum_{p}^{N_{plaq} } u_p,
\label{utp}
\end{equation}
where each plaquette consists of a square with four sites enclosing an elementary cell of 
the lattice, where we have taken into account that each pair interaction is counted in two different plaquettes.
In Table \ref{table.maps} we collect the energies for representative plaquette configurations of both models.
\begin{table}
\begin{tabular}{|cc|c|c|c|c|}
\hline \hline
& Plaquette &  
$\begin{array}{cc} + & + \\ + & + \end{array} $  & 
$\begin{array}{cc} + & + \\ + & - \end{array} $  &   
$\begin{array}{cc} + & + \\ - & - \end{array} $  &   
$\begin{array}{cc} + & - \\ - & + \end{array} $   \\
\hline
Ising &  $u_{p}/\epsilon_{I}$  & -4 & 0  & 0 &  4 \\
\hline
SARR & $u_{p}/\epsilon$  & - 2  & -1 & - 1 & 0\\ 
 \hline \hline
\end{tabular}
\caption{Plaquette interactions in the Ising and the FLL of the SARR models on the square lattice.}
\label{table.maps}
\end{table}
The mapping between the two models, is then for any plaquette configuration:
\begin{equation}
u_p^{Ising}/\epsilon_I = 4 u_p^{FLL}/\epsilon + 4.
\end{equation}
implying that the FLL limit of the SARR model, on the square lattice, is in the 2D Ising 
universality class.

\subsection{The zero density limit: Self-assembly}

The SARR model has two independent thermodynamic parameters, the temperature and the density of monomers. At 
high temperatures $k_BT >> \epsilon$ there is little bonding 
and the behavior of the model is similar to that of the lattice gas. At low temperatures, however, 
bonding dominates and the model behaves in a strikingly different way. Rods self-assemble and, 
at a fixed density, the average rod length increases exponentially as the temperature decreases. 
The polydisperse rods undergo an ordering transition at a density that is temperature dependent. 
The transition was calculated in Ref. \cite{Tavares2009a} using a generalized mean-field 
theory of self-assembly, where the polydisperse rods interact through Onsager-like excluded 
volume terms only. 

\begin{figure}[h]
\includegraphics[width=100mm,clip=]{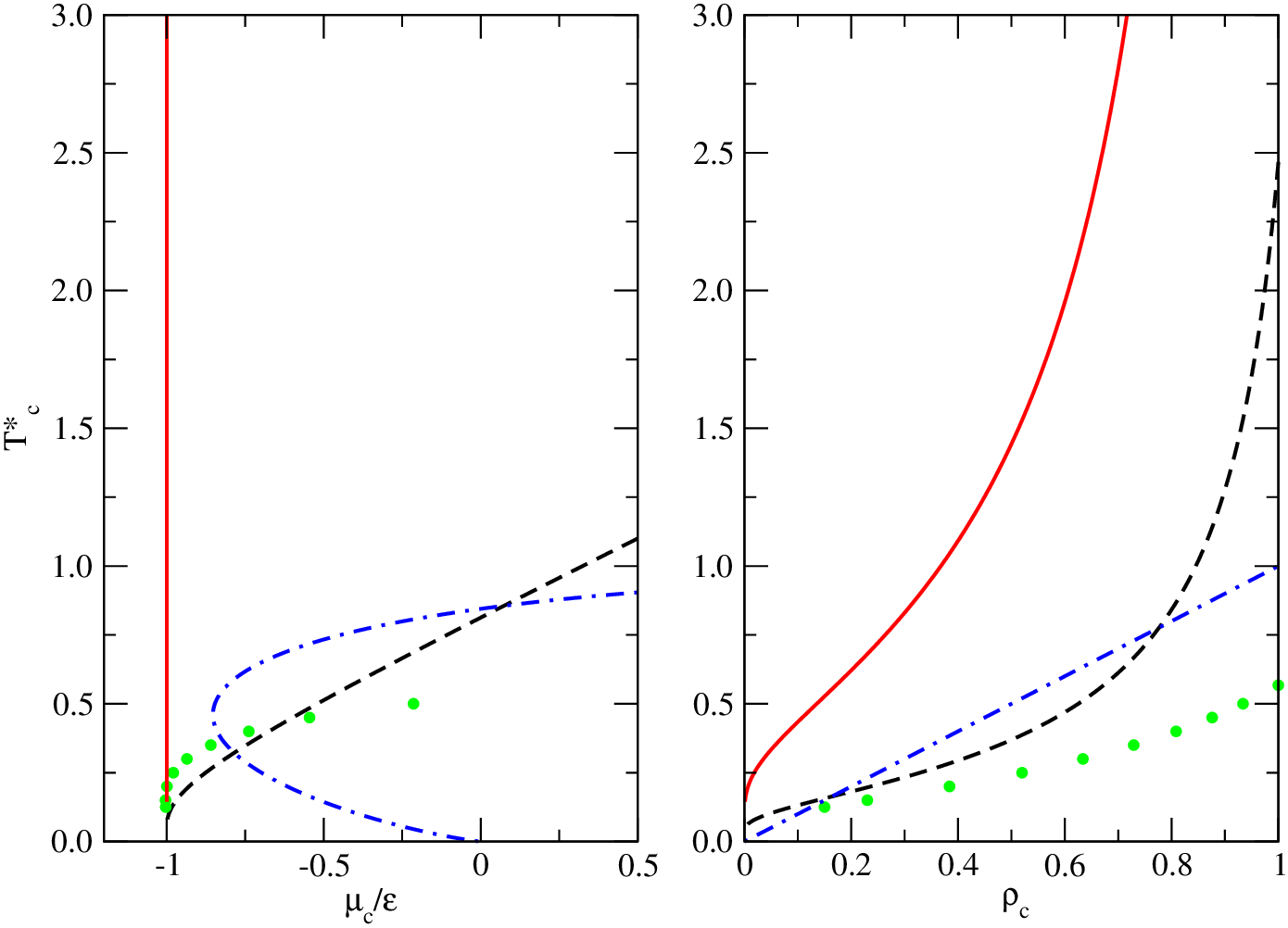}
\caption{(Color on line) Critical lines from MC simulation (points), theory of self-assembly (dashed)
and mean-field theory (dashed-dotted). Crossover line is the full line. Left panel: $\mu, T$ diagram. 
Right panel: $\rho, T$ diagram. See text for details.}
\label{criticaline}
\end{figure}

The critical line is given by\cite{Tavares2009a}:
\begin{equation}
\label{critlinerhoT}
\frac{1}{T^*_c}=\ln\left[\frac{(2-\rho_c)(2+\rho_c)}{2\rho_c^3}\right],
\end{equation}
and is plotted in Fig. \ref{criticaline}. This line is singular in the zero-density limit, where the 
average rod length diverges, signalling the self-assembly or equilibrium polymerization transition 
at zero temperature. An estimate of the crossover line, from the polymerzation transition, is obtained from 
the asymptotic relation between $T^*_c$ and $\rho_c$, at the singular point:
\begin{equation}
\label{crossovTrho}
T^*_c \sim -\frac{1}{\ln \rho_c}.
\end{equation}

Using the results for the thermodynamic potentials derived in Ref. \cite{Tavares2009a} one finds that 
the chemical potential at the critical point, $\mu_c$, is given in terms of the critical temperature 
and density, $T_c$ and $\rho_c$:
\begin{equation}
\label{critlinemuT}
\mu_c/\epsilon = T^*_c\left[\ln(2\rho_c^3)+\rho_c/2-2\ln(2-\rho_c)\right].
\end{equation}
Using the asymptotic form, Eq. (\ref{crossovTrho}), we find for the crossover line:
\begin{equation}
\label{crossovTmu}
\mu_c/\epsilon  \sim -1.
\end{equation}
This line delimits the region where the self-assembly or equilibrium polymerization fluctuations are large.
In Fig. \ref{criticaline} we plot the critical and crossover lines of the SARR model as functions of
$\mu$ and $\rho$. Note that the crossover line approaches the critical line tangentially in the
equilibrium polymerization limit, suggesting that the asymptotic scaling region of the finite density
critical point decreases rapidly as the critical temperature and density decrease.  

Finally, the internal energy on the critical line is easily calculated and we find\cite{Tavares2009a}: 
\begin{equation}
\label{intensacl}
\frac{U}{N\epsilon}=-\frac{2\rho_c}{2+\rho_c}.
\end{equation}
Note that the energy per particle increases monotonically with the (critical) density. This is plotted in 
Fig. \ref{intenergycrit} and will be discussed in the next section. 

\subsection{Intermediate densities}

A look at Table \ref{table.tc} reveals that the critical energy per particle, 
$u_c = <U/N>_c$, varies non-monotonically with the temperature. In the self-assembly limit, 
the internal energy $u^* = <U/N\epsilon>$ is a measure of the average rod length and is predicted 
to vary monotonically on the critical line, as stated above (see Ref.\cite{Tavares2009a} for details). 
Clearly, a departure from this behavior indicates the importance of the attractions between rods, 
which are short in the high-temperature regime. 

Let us assume that there is no bonding, i.e., the temperature is so high that the average rod length is
of order 1. Then, on average, each monomer interacts with its aligned neighbors, the pairs being aligned 
with the corresponding lattice bonds. The mean-field free energy becomes:
\begin{equation}
\label{freeenmf}
\beta f =\sum_{\alpha=x,y}\rho_\alpha(\ln\rho_\alpha -1) +
(1-\rho)(\ln(1-\rho)-1)-\beta \epsilon (\rho_x^2+\rho_y^2),
\end{equation}
where $\rho_x$ and $\rho_y$ are the densities of particles aligned with the $\hat x$ and $\hat y$ 
directions, respectively, and we have accounted for the entropy of the empty lattice sites. 
Given that $\rho=\rho_x+\rho_y$ and defining $\Delta=\rho_x-\rho_y$, the free energy may be written 
in terms of these variables.
The critical points are obtained by (i) calculating the field associated with $\Delta$, 
$\beta h=\frac{\partial \beta f}{\partial \Delta}$ and (ii) setting $h=0$ to obtain, implicitly, 
$\Delta(\rho,T)$. The critical line is given by:
\begin{equation}
\label{critlinerhoTmf}
T^*_c=\rho_c.
\end{equation}
and the internal energy on the critical line becomes:
\begin{equation}
\frac{U}{N\epsilon}=-\frac{\rho_c}{2}.
\label{mfinternergy}
\end{equation}

\begin{figure}[h]
\includegraphics[width=100mm,clip=]{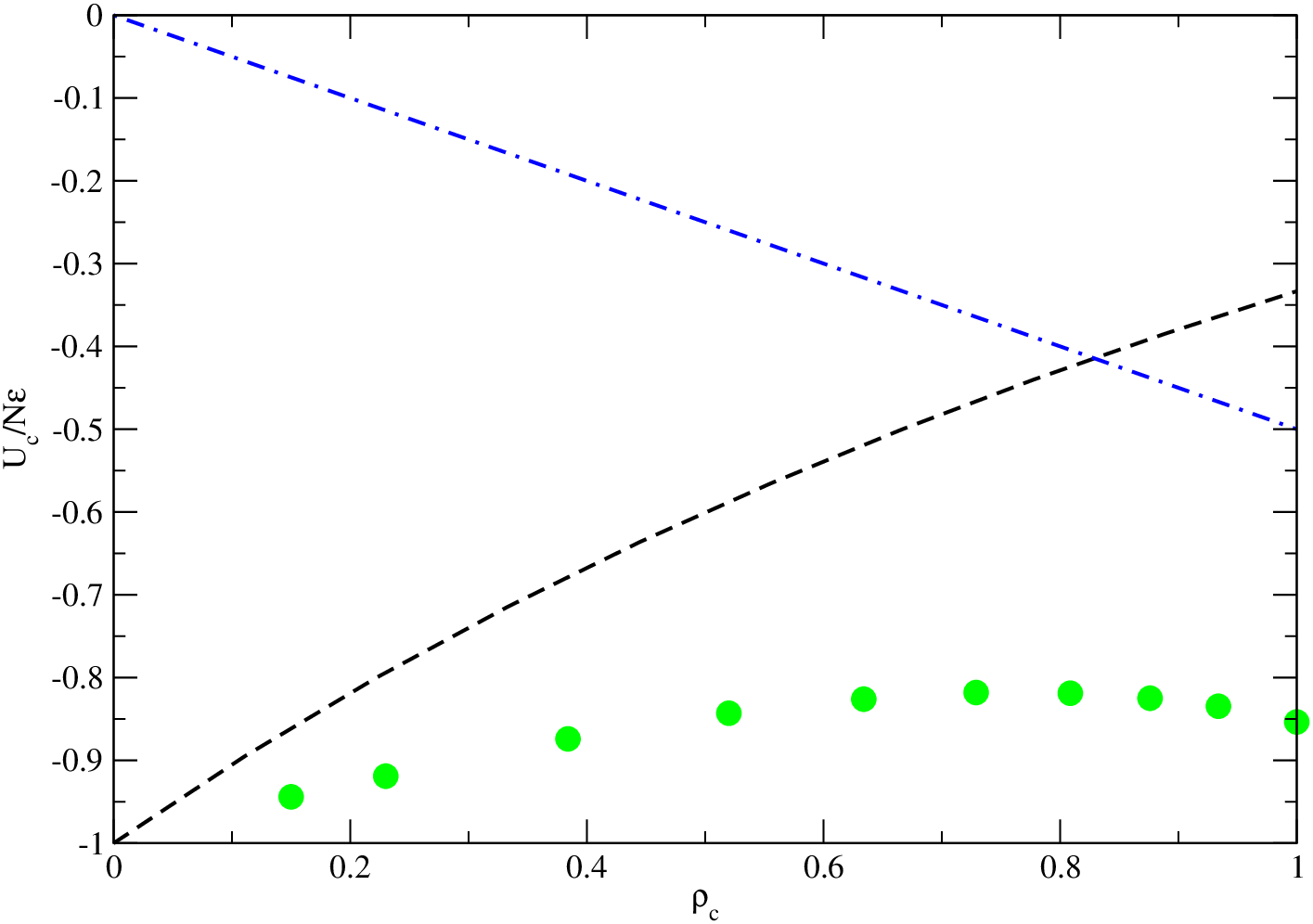}
\caption{(Color on line) Internal energy per particle on the critical line. Points: MC simulation, dashed: theory of self-assembly, 
dashed-dotted: mean-field theory. See text for details.}
\label{intenergycrit}
\end{figure}

The internal energy on the critical line is plotted in Fig. \ref{intenergycrit}.
The points are the computer simulation results while the two lines are obtained from the self-assembly and 
the high-temperature mean-field theories. The results from self-assembly, Eq. (\ref{intensacl}), are monotonically 
increasing, while those from the high temperature theory, Eq. (\ref{mfinternergy}), are monotonically decreasing.

The low density/temperature behavior is captured by the self-assembly theory \cite{Tavares2009a} while the high density/temperature 
limit is described by the mean-field theory. 
Similar remarks apply to the critical line itself, as shown in Fig. \ref{criticaline}. 

This analysis suggests that although self-assembly fluctuations become increasingly important as the density 
decreases the nature of the singularity changes at $\rho=0$ only. Nevertheless, the scaling region decreases rapidly 
in the low-density/temperature region and the true asymptotic behavior may be difficult to observe in simulations of 
reasonably sized systems.   

\section{Discussion}

The results reported in the previous sections clearly suggest that the critical behavior of the
SARR model is 2D Ising. This conclusion is supported by (i) 
the scaling behavior of the Binder cumulant for different system sizes, (ii) the system size 
dependence of the peaks of $\rho'_{\mu}(\mu)$ and (iii) the values of the critical exponent $\nu$.
This conclusion contrasts with that of L\'opez et al., and it is important to understand
the reasons for this discrepancy.
First we acknowledge that the values of $\beta/\nu$ are relatively similar for the two universality 
classes; the same may be said of $\gamma/\nu$. Thus the distinction between the two universality classes  
will have to be based on the value of the $g_4(L)$ crossing and on the value of $\nu$.
In the analysis of L\'opez et al, the use of the density as the control parameter leads to a
value of the $g_4$ crossing that differs substantially from that of the 2D 
Ising universality class. We have shown that using $\mu$ as the control parameter leads to 
a more robust scaling of $g_4$ and to a much better overall Ising scaling. 

\begin{figure}
\includegraphics[width=80mm,clip=]{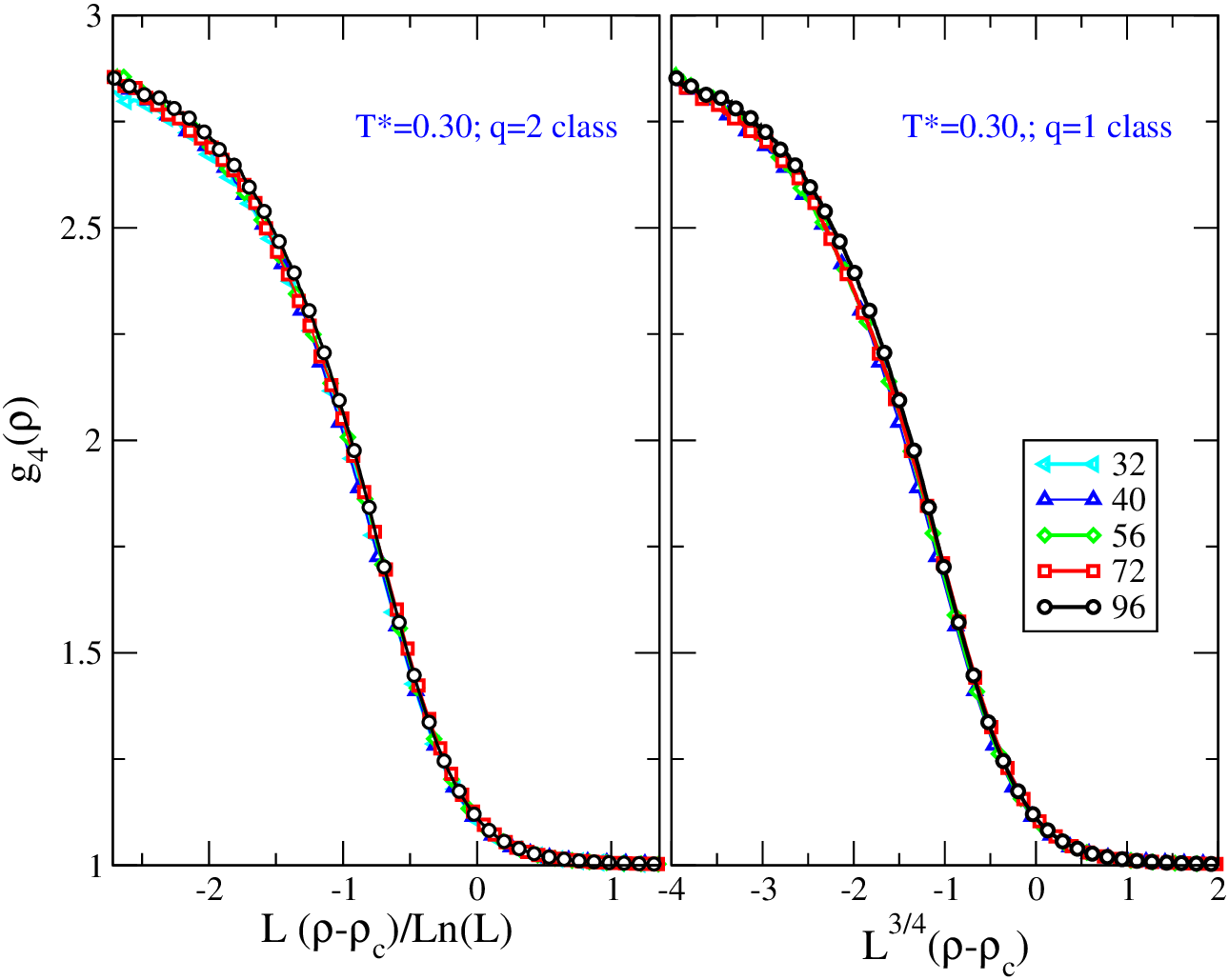}
\caption{(Color on line) Fourth order Binder cumulants for different system sizes 
as a function of the scaling densities in the $q=1$ and $q=2$ universality classes, at $T^*=0.30$.}
\label{fig.rho_scaling}
\end{figure}

Concluding that the SARR model is indeed in the Ising universality class, the question is then,
how was the value of $\nu \approx 4/3$ observed when using $\rho$ as the control parameter?
Consider a property $Q$ whose derivative with respect 
to the control parameter has a maximum in the critical region. 
Such derivative scales in the finite size region as\cite{Lopez,Landau_Binder}:
\begin{equation}
 Q'_{s} \equiv \left( \frac{ \partial Q(L)}{\partial s}\right)^{max} = a L^{1/\nu} ( 1 + b L ^{-\omega} ), 
\label{dqdx}
\end{equation}
where $s$ represents either the density or the chemical potential. 
In the scaling region $Q'_{\mu}$, and $Q'_{\rho}$ are related by:
\begin{equation}
Q'_{\rho} \approx Q'_{\mu} \left( \frac {\partial \mu }{\partial \rho} \right) \sim  \frac{L}{\ln L},
\end{equation}
where we used $\nu=1$ (Q2UC) and the scaling relation given in Eq. (\ref{ddq2}). 
We suspect that the presence of $\ln L$ in the scaling of $Q'_{\rho}$, raises the value of 
the effective critical exponent $\nu$. In particular, for the range $60 \le L \le 120$ used by L\'opez et al. 
\cite{Lopez} the ratio $L/\ln L$ is well described by: $L /\ln L \simeq a L^{1/\nu'}$, with 
$\nu' \simeq 1.291$, close to the value $\nu = 4/3$ of the Q1UC. This is illustrated in Fig. 
\ref{fig.rho_scaling} where it is clear that the finite-size scaling of $g_4(\rho,L)$ 
is well described using both $x=L^{-3/4}(\rho-\rho_c)$ and $x = (\rho - \rho_c) L /\ln L$ as 
the scaling density.
To conclude, we have shown that the criticality of the SARR model is 2D Ising. Nevertheless, as the 
temperature decreases deviations from the Ising scaling laws increase, and larger system sizes are needed to 
obtain accurate estimates of the critical exponents. 
This can be understood in terms of the self-assembly fluctuations that occur closer to the critical line as the 
density and temperature decrease. In addition, the use of PBC may enhance this finite-size effect, through the 
percolation of 'periodic' rods.

Finally, we note that Milchev and Landau \cite{Milchev} analysed the critical behavior of a flexible 
self-assembling rod model in 2D. They report a continuous transition in the $T,\mu$ space ending at a 
tricritical point, at finite density, and critical exponents on the continuous portion of the two-phase boundary 
in the 2D Ising class. Their model is richer than ours but the nature of the continuous portion of the phase boundary 
is likely to be the same. The connection between these two models as well as extensions to 3D will be left 
for future work. 

\acknowledgments
NGA gratefully acknowledges the support from the Direcci\'on General de Investigaci\'on 
Cient\'{\i}fica  y T\'ecnica under Grants Nos. MAT2007-65711-C04-04 and FIS2010-15502, and from the
Direcci\'on General de Universidades e Investigaci\'on de la Comunidad
de Madrid under Grant No. S2009/ESP-1691 and Program MODELICO-CM. MMTG and JMT acknowledge 
financial support from the Portuguese Foundation for Science and Technology (FCT) under 
Contracts nos.\ POCTI/ISFL/2/618 and PTDC/FIS/098254/2008.

\end{document}